\documentclass[aps,preprint, superscriptaddress,nofootinbib]{revtex4}
\usepackage{amsmath}
\usepackage{graphicx}
\usepackage{subfigure}
\usepackage{amssymb}
\usepackage{xcolor}
\usepackage{multirow}
\usepackage{cancel}
\usepackage{color}
\usepackage{epsfig}
\usepackage{ulem}
\usepackage{listings}
\usepackage{xcolor}
\usepackage{float}
\usepackage{slashed}

\usepackage{tikz}
\usepackage{pgffor}							

\usepackage[colorlinks,citecolor=blue]{hyperref}
\usepackage{amsmath}
\usepackage{wrapfig}

\newcommand{\be}{\begin{equation}}
\newcommand{\ee}{\end{equation}}
\newcommand{\beq}{\begin{equation}}
\newcommand{\eeq}{\end{equation}}
\newcommand{\bea}{\begin{eqnarray}}
\newcommand{\eea}{\end{eqnarray}}
\newcommand{\besp}{\begin{equation}\begin{split}}
\newcommand{\eesp}{\end{split}\end{equation}}

\newcommand{\Dfbd}{\mathord{\buildrel{\lower3pt\hbox{$\scriptscriptstyle\leftrightarrow$}}\over {D}_{\mu}}}

\def\0{\textbf{0}}
\def\1{\textbf{1}}
\def\2{\textbf{2}}
\def\3{\textbf{3}}
\def\4{\textbf{4}}
\def\5{\textbf{5}}
\def\6{\textbf{6}}
\def\7{\textbf{7}}
\def\8{\textbf{8}}
\def\9{\textbf{9}}

\begin{document}

\title{Direct detection of cosmic ray-boosted puffy dark matter}

\author{Wenyu Wang}
\email{wywang@bjut.edu.cn}
\affiliation{Faculty of Science, Beijing University of Technology, Beijing, China}

\author{Wu-Long Xu}
\email{wlxu@emails.bjut.edu.cn}
\affiliation{Faculty of Science, Beijing University of Technology, Beijing, China}

\author{Jin Min Yang}
\email{jmyang@itp.ac.cn}
\affiliation{CAS Key Laboratory of Theoretical Physics, Institute of Theoretical Physics, Chinese Academy of Sciences, Beijing 100190, P. R. China}
\affiliation{ School of Physical Sciences, University of Chinese Academy of Sciences,  Beijing 100049, P. R. China}

\author{Rui Zhu}
\email{zhurui@itp.ac.cn}
\affiliation{CAS Key Laboratory of Theoretical Physics, Institute of Theoretical Physics, Chinese Academy of Sciences, Beijing 100190, P. R. China}
\affiliation{ School of Physical Sciences, University of Chinese Academy of Sciences,  Beijing 100049, P. R. China}


\begin{abstract}
For the light relativistic dark matter (DM) boosted by high energy cosmic ray, its scattering cross section with the nucleon is sensitively dependent on the momentum-transfer and such an dependence is caused by the mediator in the scattering. For puffy DM particle with a size,  the momentum-transfer dependence can also arise from the DM radius effect. All these momentum-transfer dependences should be considered. 
In this note we study the direct detection limits on the cosmic ray-boosted puffy DM for a simplified model with a light mediator.
For comparison, we first re-derive the direct detection limits on the cosmic ray-boosted point-like DM. 
We display the limits on various planes of parameters and find that the
limits for the cosmic ray-boosted puffy DM are stronger than for the point-like DM.

\end{abstract}

\maketitle

\tableofcontents

\section{Introduction}\label{sec1}
 As a crucial component of our universe, dark matter (DM) has significant influence on the evolution of the universe and the formation of large-scale structures~\cite{Springel:2006vs,Bahcall:1999xn}.  So far all evidence for the existences of DM have primarily been based on gravitational interaction, such as the  rotation curves of galaxies, the collision of bullet clusters and gravitational lensing effects~\cite{Ostriker:1973uit,Randall:2008ppe,Clowe:2003tk}. As a fundamental particle, the nature of DM particle remains a great mystery to the field of particle physics, such as its spin, mass and size \cite{Planck:2018vyg}, although the state-of-the-art technologies have been being used to detect the DM particle, especially its most popular candidate called the Weakly Interacting Massive Particle (WIMP) \cite{Lee:1977ua,Jungman:1995df,Buchmueller:2017qhf}.  

In the direct detection of DM, the traditional experiments have good sensitivities for DM mass above $1 ~\rm GeV$ \cite{GAMBIT:2017zdo}, while below $1 ~\rm GeV$ the sensitivity of detection is almost lost since the maximal non-relativity DM-nuclear recoil energy drops below the threshold of the detectors~\cite{XENON:2018voc,LUX:2017ree,PandaX-II:2018xpz,Elor:2021swj,}.  Recently, some  experiments start focusing on inelastic scattering processes involving sub-$\rm GeV$ DM, such as the DM-electron scattering, the nuclear recoils through the Migdal effect and the DM-phonon scattering \cite{Guo:2021imc,XENON:2019gfn,Essig:2017kqs,Baxter:2019pnz,Wu:2022jln,Wang:2021oha,Campbell-Deem:2022fqm,Flambaum:2020xxo,Wang:2023xgm,Li:2022acp,Gu:2022vgb}.  Furthermore, these experiments are exploring a variety of targets including noble gas atoms, semiconductors, Dirac materials and superconductors \cite{Knapen:2020aky,Liang:2022xbu,Hochberg:2016ajh,Hochberg:2015pha,Coskuner:2019odd,Kurinsky:2019pgb}. 
 
On the other hand, the direct detection of sub-$\rm GeV$ DM can become feasible, considering that the high energy galactic cosmic rays can collide with the cold sub-$\rm GeV$ DM and generate a (semi-)relativistic DM component of cosmic rays, called the CRDM flux \cite{Bringmann:2018cvk,Bondarenko:2019vrb,Wang:2021nbf,Xia:2022tid,Xia:2021vbz,Xia:2020apm,PandaX-II:2021kai,Ge:2020yuf,Bell:2021xff,Feng:2021hyz,Cappiello:2018hsu,Alvey:2019zaa,Guo:2020oum,Maity:2022exk,Ema:2018bih,PandaX:2023tfq,Su:2022wpj,Su:2020zny,Bramante:2021dyx}.  In this case, the relativistic scattering cross section between DM and ordinary matter is dependent on the momentum-transfer closely, particularly when the mediator mass is light \cite{Dent:2019krz}.  Thus the complete propagator should be considered in the calculation of cross section.  In fact,  from $1/(q^2 +m_{\phi}^2) \propto 1/(1+q^2/m_{\phi}^2)$ we see that one source of momentum-transfer dependence in the cross section comes from the mediator. Furthermore, for puffy DM with a finite size, another source of momentum-transfer dependence in the cross section arises from its radius effect, similar to the study for addressing the small-scale structure anomalies~\cite{Chu:2018faw,Wang:2021tjf,Wang:2023xii}.  For the direct detection of cosmic ray-boosted puffy DM, these two sources of momentum-transfer dependence need to be jointly considered, which is the aim of this work. The obtained limits will also provide information on the DM radius, which is crucial in revealing the nature of DM.

This work is arranged as follows. In Sec. II,  a simplified model for puffy DM is presented.  In Sec. III,  the full processes of cosmic ray-boosted puffy dark matter are described.  In Sec. IV, some numerical results about the mediators and the finite-size effects are shown.  Sec. V gives our conclusions.

\section{A simplified model with mediator for puffy dark matter}\label{sec2}
Light mediators play a crucial role in DM direct detection and collider searches, as the recoils of targets from momentum transfer are significantly influenced by the mediators. From puffy DM, the effects of DM size, which provide an additional source of momentum-transfer dependence in the DM-nucleon scattering cross section, must be fully considered to account for all factors influencing the momentum transfer \cite{Hardy:2015boa}.

When the DM particle is point-like, a simplified model with scalar $\phi$ or vector $V_{\mu}$ as the mediator between DM and nucleon has a Lagrangian  
\be\label{langrage}
\begin{split}
{\cal L}_{int} \supset &g_{\chi S}\phi \bar{\chi} \chi+g_{NS}\phi \bar{N} N\\
&+g_{D}V_{\mu} \bar{\chi} \gamma^{\mu}\chi+g_{\epsilon}V_{\mu} \bar{N} \gamma^{\mu}N,
\end{split}
\ee 
where  $\chi$ is the DM, $N$ denotes proton or neutron, and $g_{\chi s}$, $g_{N s}$, $g_{D}$ and $g_{\epsilon}$ are the coupling constants.  In general, a cold DM particle can be approximately considered at rest. However, when it is boosted by cosmic rays, its velocity can be relativistic. The differential scattering cross section between a CR particle and a point-like DM particle can be calculated as 
\bea\label{sec}
\left(\frac{d\sigma_{\chi N}}{dT_{\chi}}\right)_{\rm{ S0,CR}}&=&\frac{A^2g_{Ns}^{2}g_{\chi s}^{2}F^{2}(q^{2})(2m_{\chi}+T_{\chi})(2m_{N}^{2}+m_{\chi}T_{\chi})}{8\pi T_{i}(T_{i}+2m_{N})(m_{\phi}^{2}+2m_{\chi}T_{\chi})^{2}}\\
\left(\frac{d\sigma_{\chi N}}{dT_{\chi}}\right)_{\rm {V0,CR}}&=&\frac{A^2g_{D}^{2}g_{\epsilon}^{2}F^{2}(q^{2}) \{2m_{\chi}(m_{N}+T_{i})^{2}-T_{\chi}[(m_{N}+m_{\chi})^{2}+2m_{\chi}T_{i}]+m_{\chi}T_{\chi}^{2}\}}{4\pi T_{i}(T_{i}+2m_{N})(m_{V}^{2}+2m_{\chi}T_{\chi})^{2}} \label{secc}
\eea
where the subscript 0 means point-like DM particle with zero radius, $T_\chi$ and $T_i$ denote respectively the kinetic energies of the 
outgoing DM particle and the incoming CR particle (proton or helium), $A$ is the number of nucleons inside a CR particle, and $F(q)$ takes a  dipole form for proton~\cite{Bringmann:2018cvk}. 

Puffy DM was proposed \cite{Chu:2018faw} to realize the self-interacting DM scenario to solve the small scale structure anomalies. Note that the literature of  constraints from indirect detection or relic density on the puffy dark matter model have not been found.  So far the studies on the puffy DM have been focusing on the self-interaction to try to solve the small-scale problems. And for the calculation of DM-nucleon cross section, a correct relic density of DM is assumed.
For the DM-nucleus scattering, when the momentum-transfer is much smaller than the inverse size of the nucleus, the scattering is considered as a coherent scattering \cite{Engel:1991wq}. Instead, when the momentum-transfer is sizable and the spin-independent scattering loses the coherence, the internal structure of the nucleus can be probed and we use a form factor for the nucleus which is related to the structure of the nucleus~\cite{Feldstein:2009tr}.  Similarly, if the DM particle has a finite size,  the DM-nucleus scattering cross section can be simplified as a product of the conventional DM-nucleus cross section with an additional form factor of DM~\cite{Chu:2018faw,Hardy:2015boa}.  In this work, we use such a form-factor approach and consider a dipole shape form factor given by 
\be\label{eq1}
F_{\rm DM}(q^2)=\frac{1}{(1+r_{\rm DM}^2q^2)^2} ~ ,
\ee
 where $r_{\rm DM}$ is the characteristic scale of the puffy DM particle and $q$ is the  momentum-transfer in the DM-nucleus scattering. Accordingly, the differential scattering cross section between a CR particle and a DM particle is 
 \be\label{differ}
 \left(\frac{d\sigma_{\chi N}}{dT_{\chi}}\right)_{\rm puffy,S(V),CR}=F_{\rm DM}^2(q^2)\left(\frac{d\sigma_{\chi N}}{dT_{\chi}}\right)_{\rm S0(V0),CR}
 \ee
In fact, as the radius of the puffy DM particle increases, the cross section becomes obviously suppressed, even for a light mediator. So, for a sizable radius, such a size effect may be crucial in the DM scattering processes  (note that the strong interaction for this puffy particle scattering is not considered in our work). 

\section{Detecting cosmic ray-boosted puffy dark matter}\label{sec3}
In this section,  the relevant scattering processes of the cosmic ray-boosted puffy dark matter (CRPDM) are described. In fact, all these processes are also involved in detecting the CRDM~\cite{Bringmann:2018cvk}.
These scattering processes include: the puffy DM particles are scattered and boosted by high energy CR,  the CRPDM is attenuated by the dense matter of the earth, the CRPDM scatters off the nuclei in the detector. Considering the size of a puffy DM particle, it is necessary to take into account the momentum-transfer arising from both the mediator effect and the DM radius effect in the scattering processes. 

{\bf (1) Puffy DM boosted by high energy CR:}  When a puffy DM particle is initially at rest and then boosted by high-energy CR, the resulting flux of the relativistic puffy DM can be expressed as
\be\label{flux}
\frac{d\Phi_{\chi}}{dT_{\chi}}=D_{\rm eff}\frac{\rho_{\chi}}{m_{\chi}}\sum_{i}\int_{T_{i}^{\rm min}}dT_{i}\left(\frac{d\sigma_{\chi i}}{dT_{\chi}}\right)_{\rm puffy, CR}\frac{d\Phi_{i}^{\rm LIS}}{dT_{i}}~,
\ee
where the effective diffusion zone parameter $D_{\rm eff}$ is set as $1~ \rm kpc$ as in \cite{Bringmann:2018cvk}, the local DM density  $\rho_{\chi}=0.3~  \rm GeV cm^{-3}$, the local interstellar (LIS) flux for protons and He nuclei is denoted as $d\Phi_i^{\rm LIS}/dT_i$ \cite{Boschini:2017fxq,DellaTorre:2016jjf}, and $T_i^{\rm min}$ is the minimal incoming CR energy  for the boosted DM energy $T_{\chi}$ given by
\be
T_i^{\rm min}=\left(\frac{T_{\chi}}{2}-m_i\right)\left[ 1\pm\sqrt{1+\frac{2T_{\chi}}{m_{\chi}}\frac{(m_i+m_{\chi})^2}{(2m_i-T_{\chi})^2}} ~\right].  
\ee

{\bf (2) CRPDM attenuated by dense matter of the earth:}  Before the relativistic CRPDM reaches the detector, it undergoes significant attenuation as it travels from the top of the atmosphere to the location of the detector.  The degradation of DM energy due to the dense matter of the earth can be estimated numerically \cite{Starkman:1990nj,Emken:2018run}
\be \label{att1}
\frac{dT_{\chi}^{z}}{dz}=-\sum_{N}n_{N}\int_{0}^{T_{N}^{max}}\frac{d\sigma_{\chi N}}{dT_{N}}T_{N}dT_{N}~,
\ee 
where $T_{\chi}^z$ denotes the DM energy located at the $z$-depth from the top of the atmosphere of the earth,  $n_N$  is the average nuclei densities of the earth's elements provided by {\bf DarkSUSY 6} \cite{Bringmann:2018lay}, $T_N$  is the recoil energy of nucleus $N$, and  the differential cross section $d\sigma_{\chi N}/dT_N$ is calculated at the $z$-depth.  The attenuated CRPDM flux located at the $z$-depth can be obtained via the primordial flux $d\Phi_{\chi}/dT_{\chi}$ 
\be \label{att2}
\frac{d\Phi_{\chi}}{dT_{\chi}^{z}}=\left(\frac{dT_{\chi}}{dT_{\chi}^{z}}\right)\frac{d\Phi_{\chi}}{dT_{\chi}} ~.
\ee

{\bf (3) CRPDM scattering in detectors:}  If the attenuated CRPDM retains enough energy, it may produce a signal that can be detected directly at the detector.   The recoil rate per target particle mass at the recoil energy range $T_1<T_N<T_2$ is given by 
\be\label{rate}
R=\int_{T_{1}}^{T_{2}}\frac{1}{m_{N}}dT_{N}\int_{T_{\chi}^{z,min}}^{\infty}dT_{\chi}^{z}\frac{d\sigma_{\chi N}}{dT_{N}}\frac{d\Phi_{\chi}}{dT_{\chi}^{z}}~.
\ee
 The differential cross section can be obtained from Eqs. (\ref{sec}, \ref{secc}, \ref{differ})  via the substitutions $m_{\chi}\leftrightarrow m_N$, $T_i \rightarrow T_{\chi}$ and $T_{\chi} \rightarrow T_N$.  The differential scattering cross section between a  target particle and a puffy DM particle can be calculated as 
 \be\label{differ1}
 \left(\frac{d\sigma_{\chi N}}{dT_{N}}\right)_{\rm puffy,S(V),CR}=F_{\rm DM}^2(q^2)\left(\frac{d\sigma_{\chi N}}{dT_{N}}\right)_{\rm S0(V0),CR},
 \ee
\bea\label{seccc}
\left(\frac{d\sigma_{\chi N}}{dT_{N}}\right)_{{\rm {S0,CR}}}&=&\frac{A^{2}g_{Ns}^{2}g_{\chi s}^{2}F^{2}(q^{2})(2m_{N}+T_{N})(2m_{\chi}^{2}+m_{N}T_{N})}{8\pi T_{\chi}(T_{\chi}+2m_{N})(m_{\phi}^{2}+2m_{N}T_{N})^{2}}\\
\left(\frac{d\sigma_{\chi N}}{dT_{N}}\right)_{{\rm {V0,CR}}}&=&\frac{A^{2}g_{D}^{2}g_{\epsilon}^{2}F^{2}(q^{2})\{2m_{N}(m_{\chi}+T_{\chi})^{2}-T_{N}[(m_{\chi}+m_{N})^{2}+2m_{N}T_{\chi}]+m_{N}T_{N}^{2}\}}{4\pi T_{\chi}(T_{\chi}+2m_{\chi})(m_{V}^{2}+2m_{N}T_{N})^{2}}. \label{secccc}
\eea
Note that the three differential cross sections for the three scattering processes are expressed in different frames.  
 
\section{Numerical results and analysis}\label{sec4}
For numerical calculations, we modify the package  {\bf darksusy 6.2.2} \cite{Bringmann:2018lay} by encoding the differential cross section with two sources of momentum-transfer dependence from mediator and radius effects.
By comparing the results of the Xenon1T experiment with the theoretical predictions, the direct detection constraints can be obtained.

\begin{figure}[htbp]
	\centering
	\includegraphics[width=7.5cm]{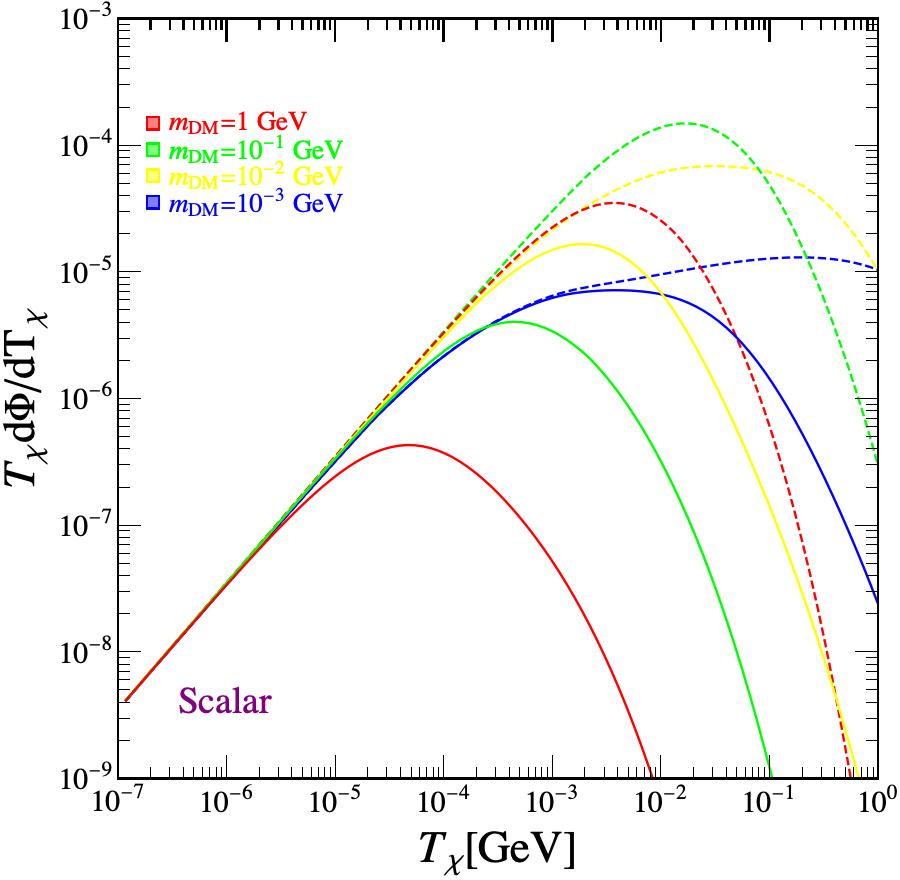}	
	\includegraphics[width=7.5cm]{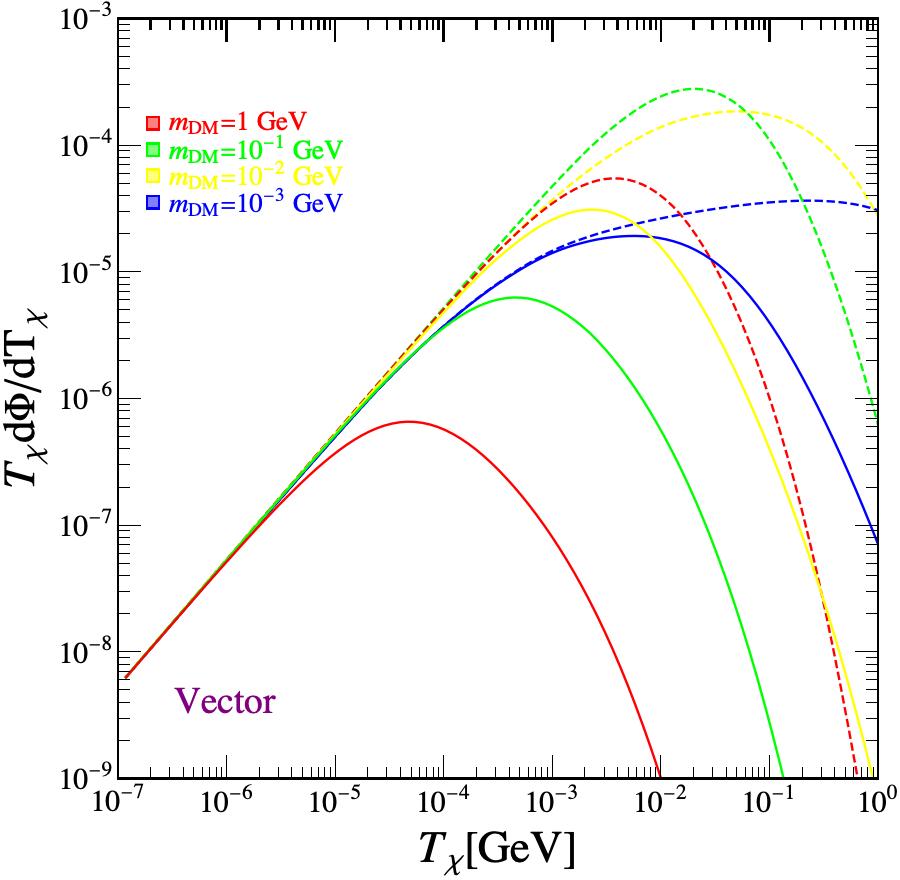}	
 \vspace{-0.4cm}
	\caption{The flux of CRPDM with $R_{\rm DM}m_{\phi}=10$ (solid curves) and CRDM (dashed curves) for different DM masses in a simplified model with a light scalar (left panel) or vector mediator (right panel) whose Lagrangian is shown in Eq.(\ref{langrage}). The mediator mass $m_{\phi}=0.1~ \rm GeV$ and the couplings are set at 1.}
	\label{fig1}
\end{figure}
First, in Fig. \ref{fig1} we present the flux of CRPDM ($R_{\rm DM}m_{\phi}=10$) and CRDM for different DM masses in the simplified model with a scalar or vector mediator  whose Lagrangian is shown in Eq.(\ref{langrage}).  We see that for different dark matter masses, both the CRDM and CRPDM fluxes can reach the (semi-)relativistic velocity range. 
For a given DM mass,  the CRPDM flux for puffy DM has a weaker strength than the CRDM flux for point-like DM. 
Also, the flux $d\Phi/dT_\chi$ (the slope of the curves in Fig. \ref{fig1}) is almost independent of the DM mass for a small $T_\chi$, contrary to the constant cross section case \cite{Bringmann:2018cvk} in which the flux decreases with the DM mass.  The  reason is  that the suppression factor $1/m_{\chi}$ in Eq. (\ref{flux}) is canceled by $m_\chi$ in the differential cross sections in Eqs. (\ref{sec}, \ref{secc}) for a small $T_\chi$. 

\begin{figure}[htbp]
	\centering
	\includegraphics[width=7cm]{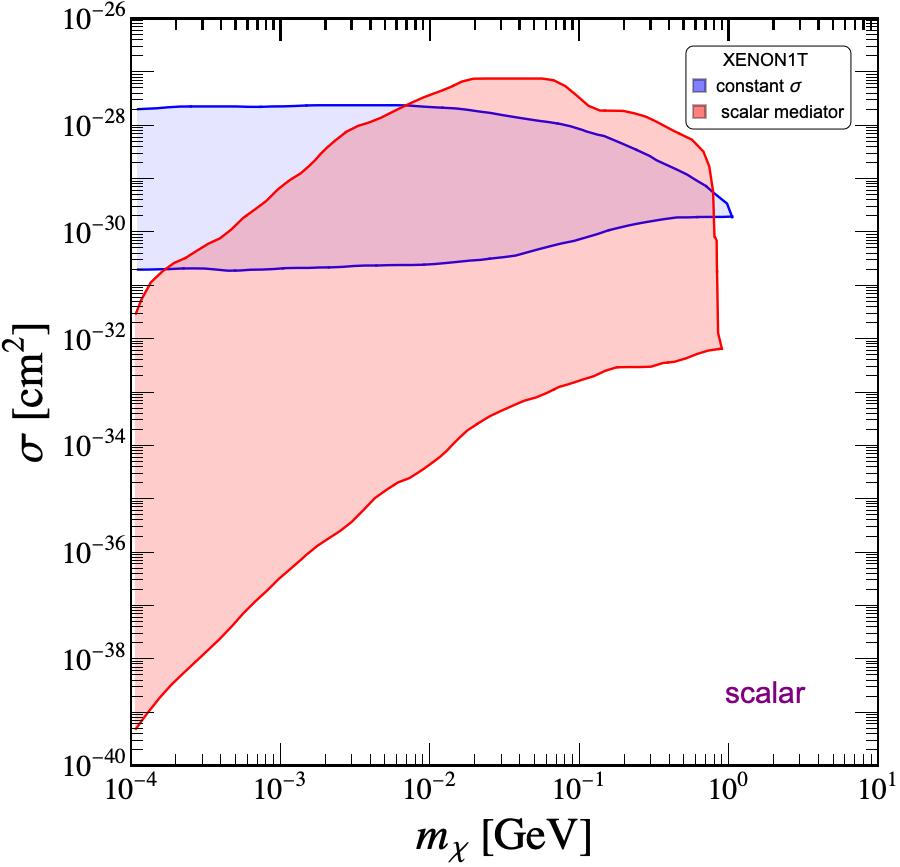}
	\includegraphics[width=7cm]{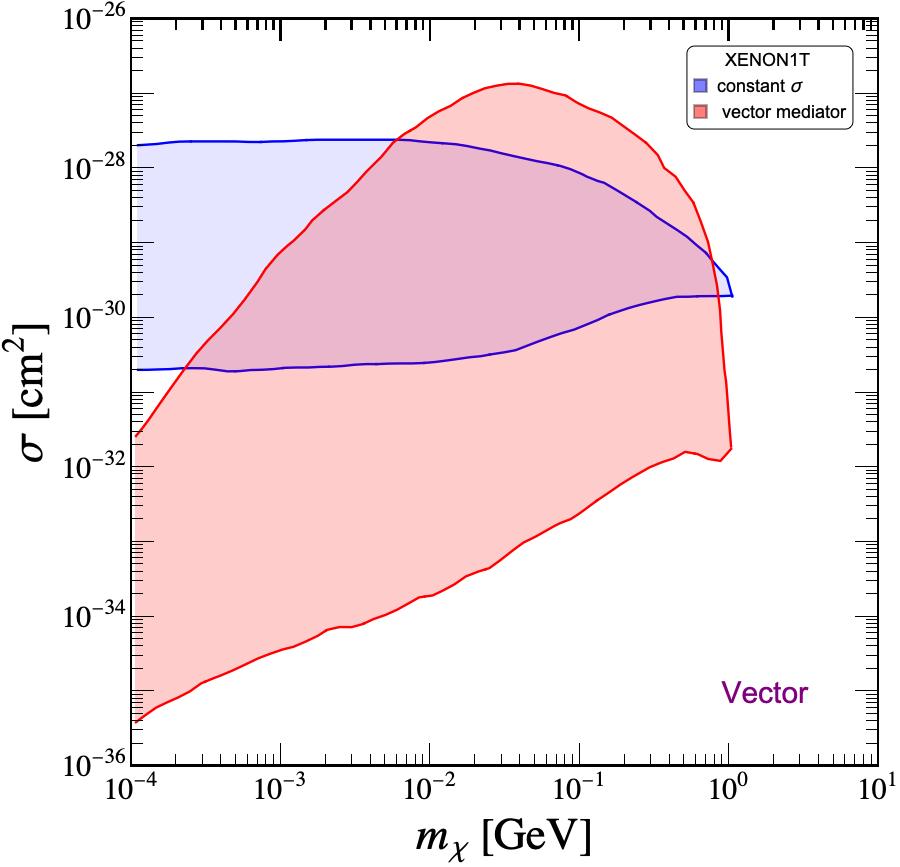}
  \vspace{-0.4cm}
	\caption{Constraints  on the spin-independent CRDM-nucleon scattering cross section from the XENON1T experiment. The light blue regions are for the constant cross section~\cite{Bringmann:2018cvk}, while the pink regions are for the simplified model with a light scalar (left panel) or vector (right panel)  whose Lagrangian is shown in Eq.(\ref{langrage}).}
	\label{fig2}
\end{figure}
\begin{figure}[htbp]
	\centering
	\includegraphics[width=7cm]{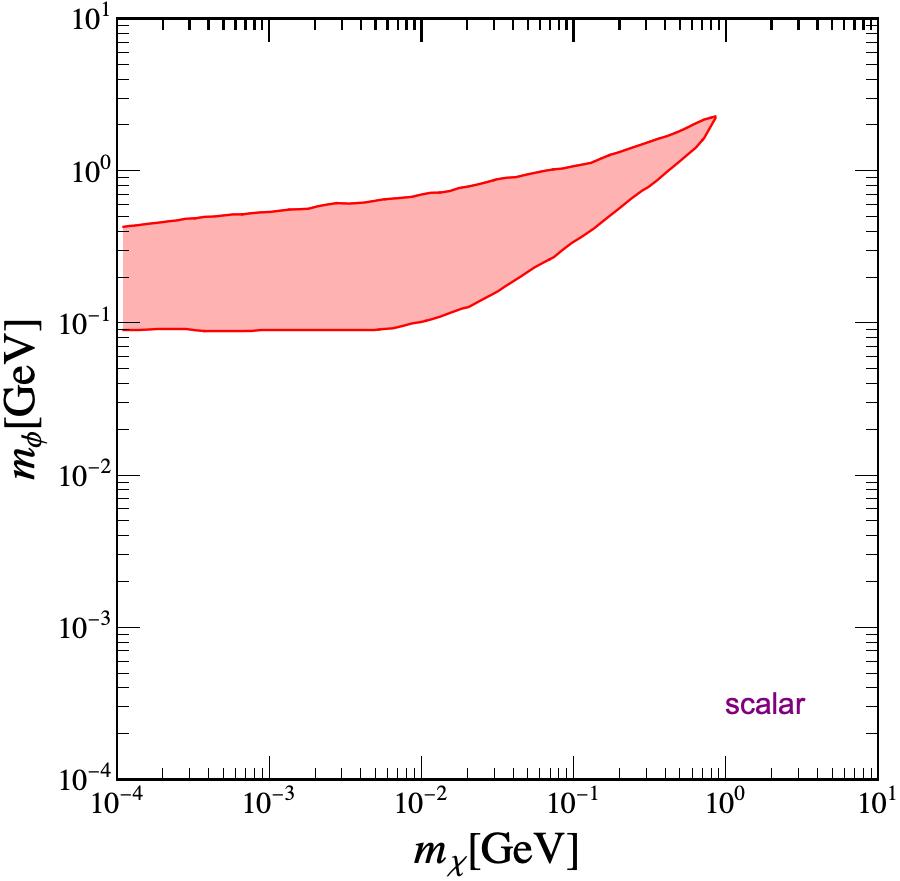}
	\includegraphics[width=7cm]{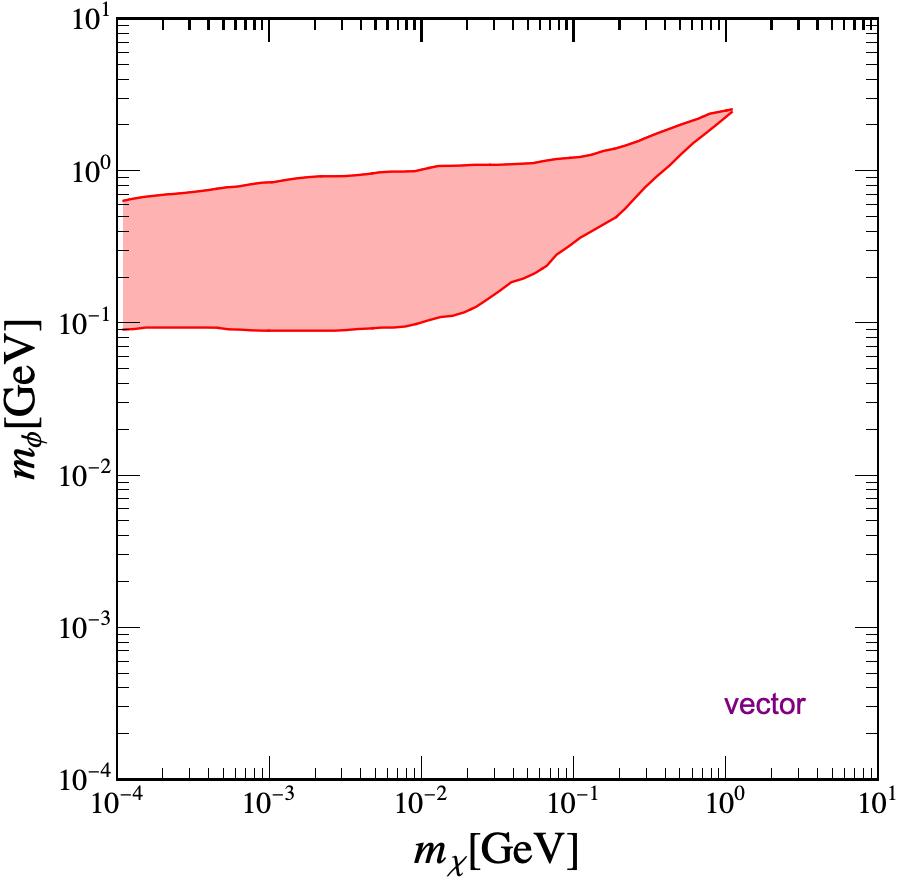}
  \vspace{-0.4cm}
	\caption{Same as  Fig. (\ref{fig2}), but showing on the plane of the CRDM mass versus the mediator mass. }
	\label{fig3}
\end{figure}

Subsequently, the constraints on the DM-nucleon spin-independent scattering cross section can be obtained from the XENON-1T experiment by analyzing the collision between the relativistic DM flux and the nucleon. We first show in Fig. (\ref{fig2}) the limits for CRDM, i.e., the point-like DM boosted by cosmic ray.  Certainly, the another more stringent constraints from direct detection, such as  LUX-ZEPLIN (LZ)'s calculated limit and sensitivities will further improve by factor $\sim 2$ compaing with the constant scattering cross section from the ~\cite{Bringmann:2018cvk} for some regions of the parameter space and upcoming XENONnT and future Darwin experiments will be sensitive to cross sections smaller by factors of $\sim 3$ and $\sim 10$ compared to the current LZ limit \cite{Maity:2022exk}.
However, from Fig. (\ref{fig2}) we see that compared with  the constant scattering cross section, stronger exclusion limits can be obtained for the DM model with a light mediator due to the momentum-transfer. A rather strong lower limit about $\mathcal{O}(10^{-40})$ is obtained in the scalar mediator case.  As analyzed in \cite{Wang:2019jtk},  the reason for the band-shape exclusion region is that the survival parameter space above the red band gives a too strong DM-nucleon cross section so the attenuation blocks CRDM to reach the detector. Of course, the survival parameter space below the band is easy to understand because it just gives a too weak CRDM-nucleon scattering cross section. 
In Fig. \ref{fig3} we demonstrate the excluded region on the plane of CRDM mass versus the mediator mass. The excluded region also takes a band shape, which means that the mediator mass is not constrained outside the band. The reason is that a too heavy mediator can severely suppress the DM-nucleon scattering cross section while  a too light mediator can lead to a too large DM-nucleon cross section so the CRDM is blocked by the strong attenuation to reach the detector.   

\begin{figure}[htbp]
	\centering
	\includegraphics[width=7cm]{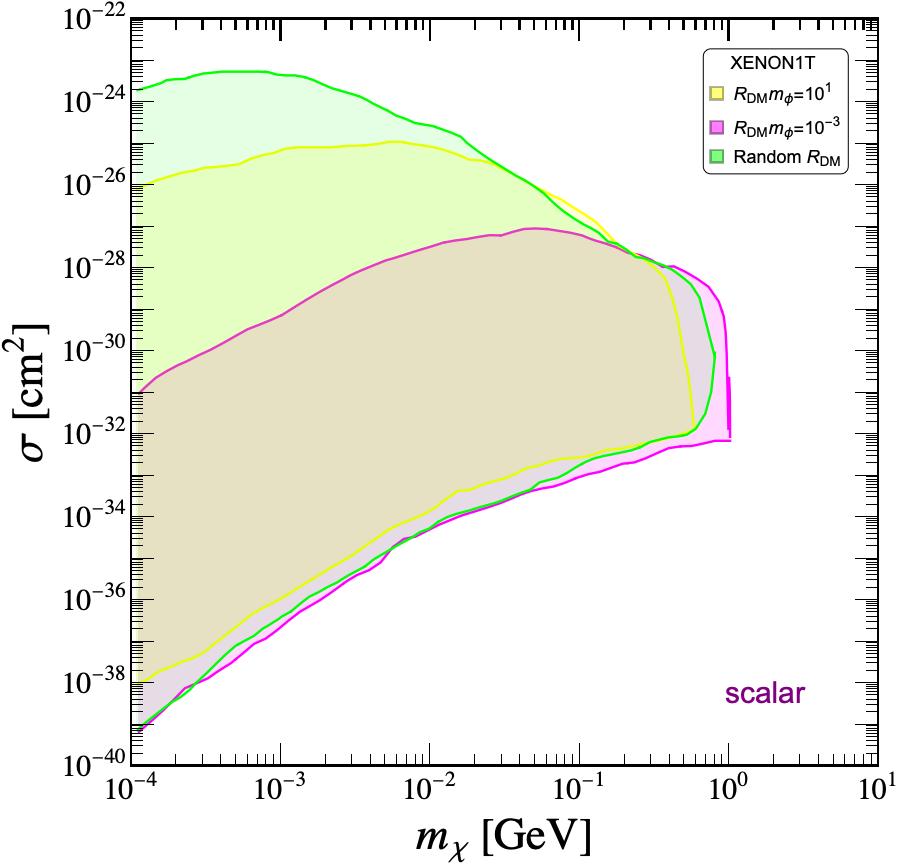}
	\includegraphics[width=7cm]{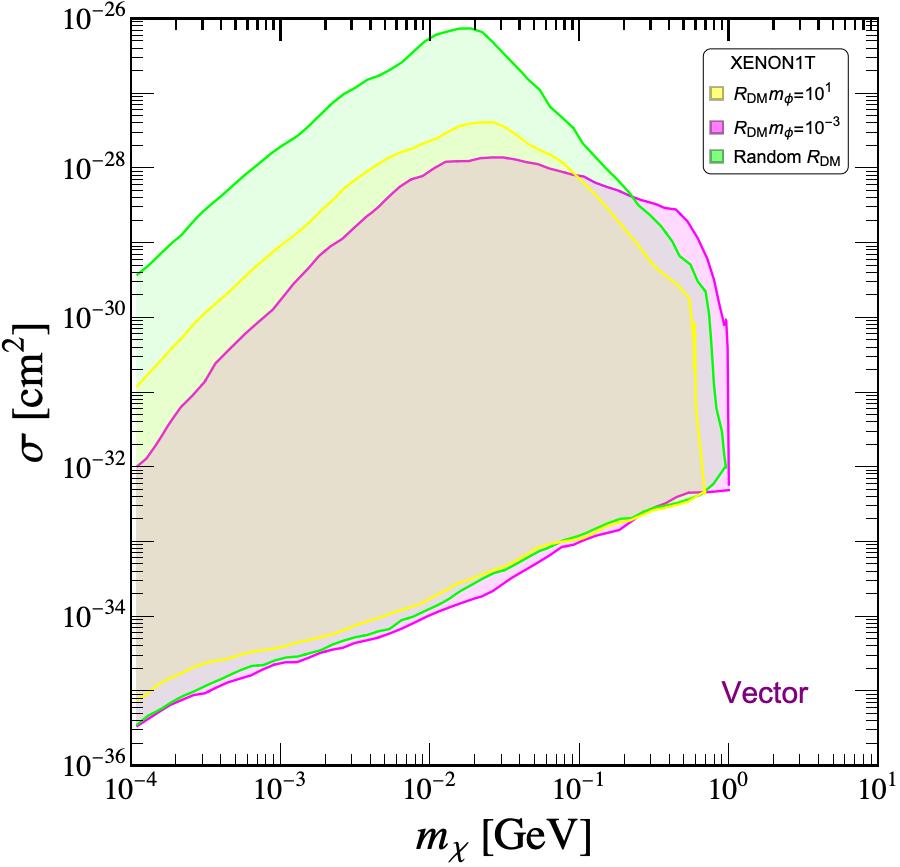}
 \vspace{-0.4cm}
	\caption{Constraints  on the spin-independent  CRPDM-nucleon scattering cross section from the XENON1T experiment. The light yellow regions and magenta regions are respectively for $R_{\rm DM}m_{\phi}=10$ and $R_{\rm DM}m_{\phi}=10^{-3}$ with the coupling constants fixed at 1.  The light green regions are for varied radius and coupling constants.}
	\label{fig4}
\end{figure}
\begin{figure}[htbp]
	\centering
	\includegraphics[width=7cm]{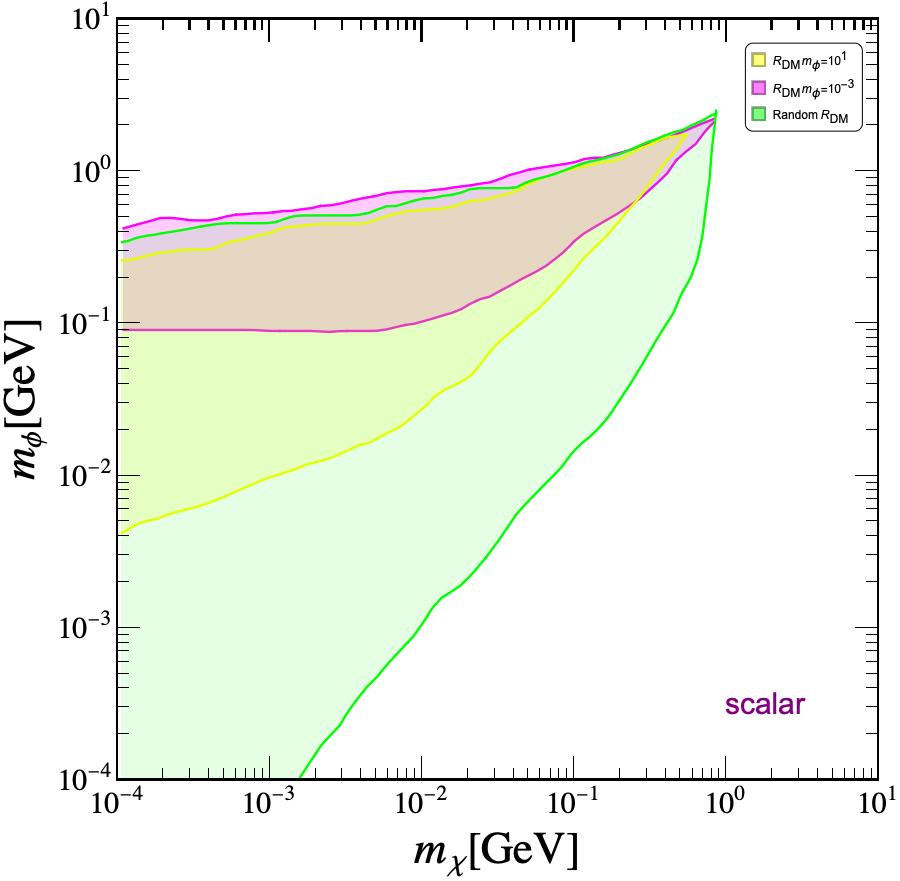}	
	\includegraphics[width=7cm]{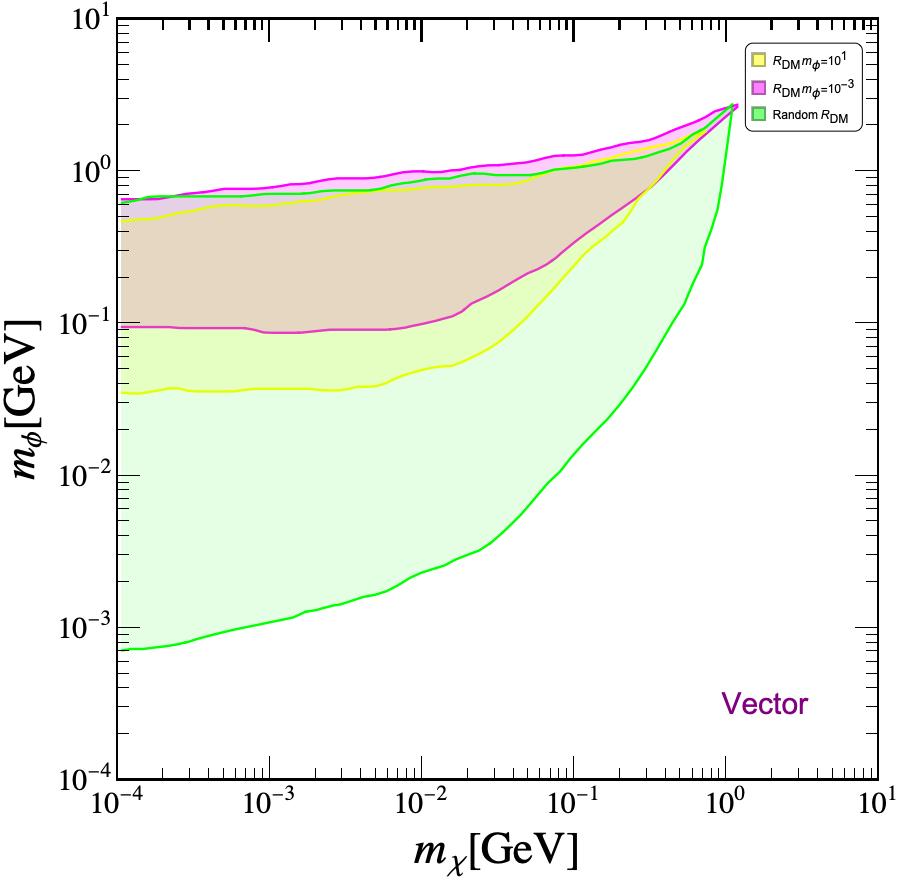}	
  \vspace{-0.4cm}
	\caption{Same as  Fig. (\ref{fig4}), but showing on the plane of DM mass versus mediator mass.}
	\label{fig5}
\end{figure}

For the CRPDM, i.e., the cosmic-ray boosted light puffy DM with a light scalar or vector mediator, 
we in Fig. \ref{fig4} show the excluded region on the plane of the spin-independent  CRPDM-nucleon scattering cross section versus the DM mass. 
It can be seen that when the mediator effect dominates (magenta regions with a radius-force range ratio $R_{\rm DM}m_{\phi}=10^{-3}$), the results are the same as for the point-like DM case with only the mediator effects shown in Fig. \ref{fig2}. However, for the radius-force range ratio $R_{\rm DM}m_{\phi}=10$, the radius effect becomes sizable and a larger excluded region with a larger cross section is obtained (yellow region). The reason is that when the radius effect is sizable, the CRPDM-nucleon scattering is suppressed and the attenuation of the CRPDM flux is weakened so that the direct detection limits become stronger. 
In Fig. \ref{fig5} we show the corresponding constraints on the plane of DM mass versus mediator mass. Again we see that the excluded region is larger for a larger radius-force range ratio.

\begin{figure}[htbp]
	\centering
	\includegraphics[width=7cm]{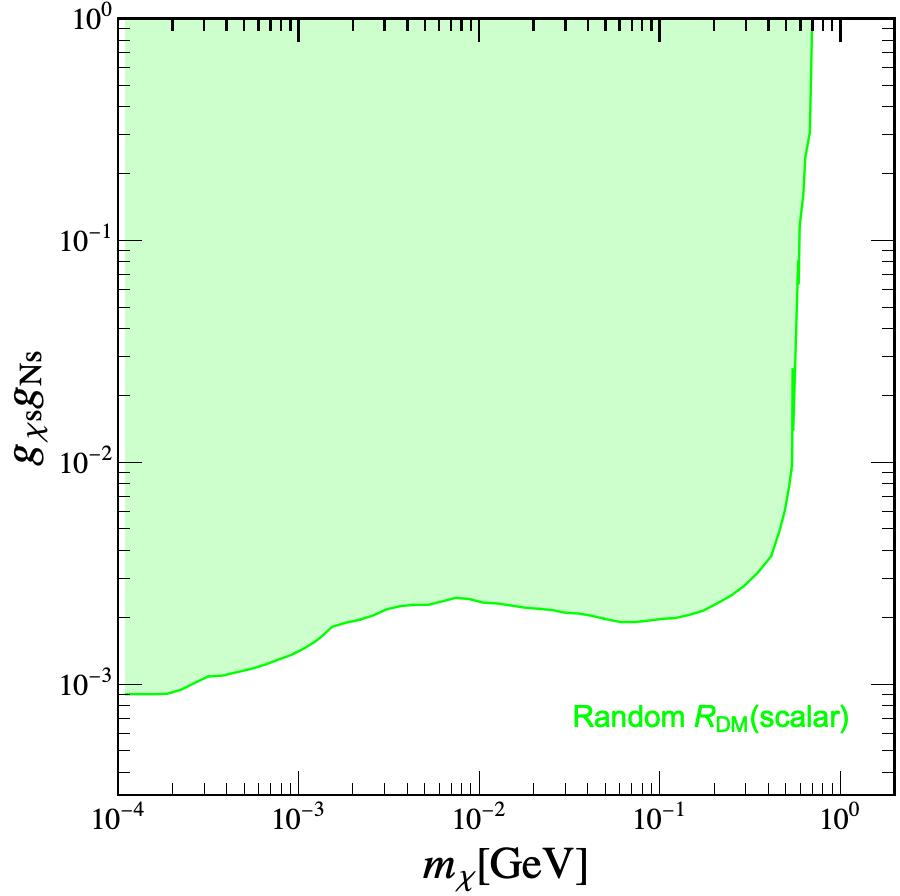}	
	\includegraphics[width=7cm]{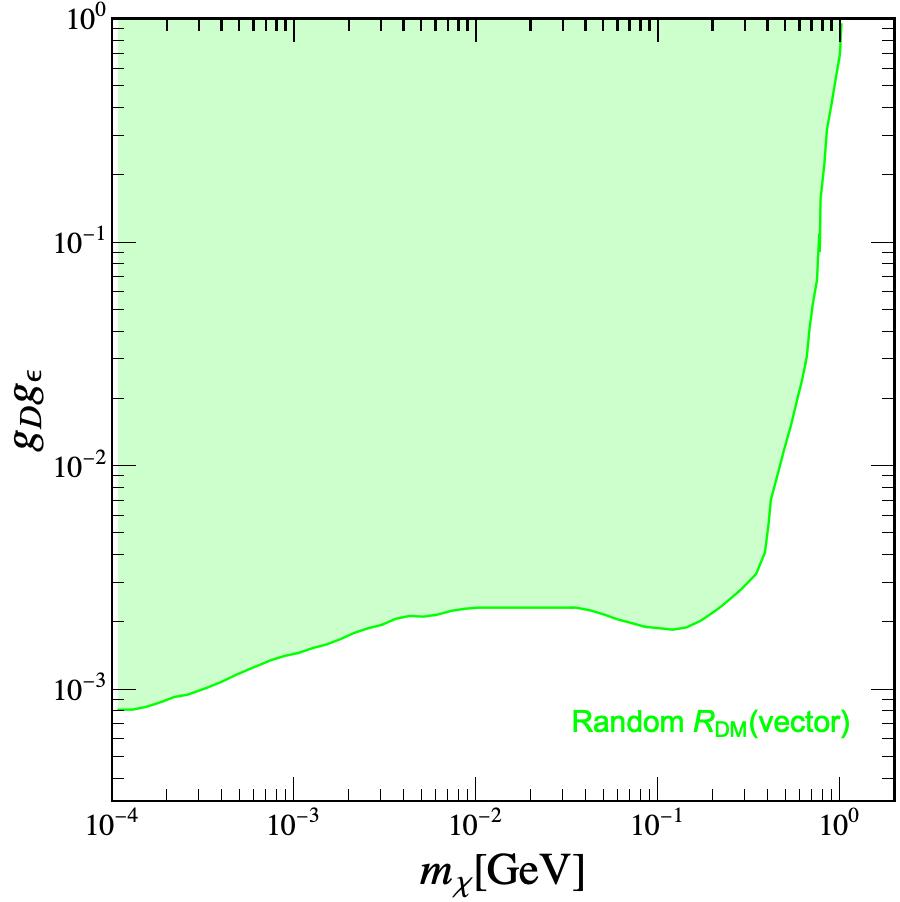}	
  \vspace{-0.4cm}
	\caption{Same as the green region in Fig. (\ref{fig4}), but showing on the plane of DM mass versus the coupling constant.}
	\label{fig6}
\end{figure}
\begin{figure}[htbp]
	\centering
	\includegraphics[width=7cm]{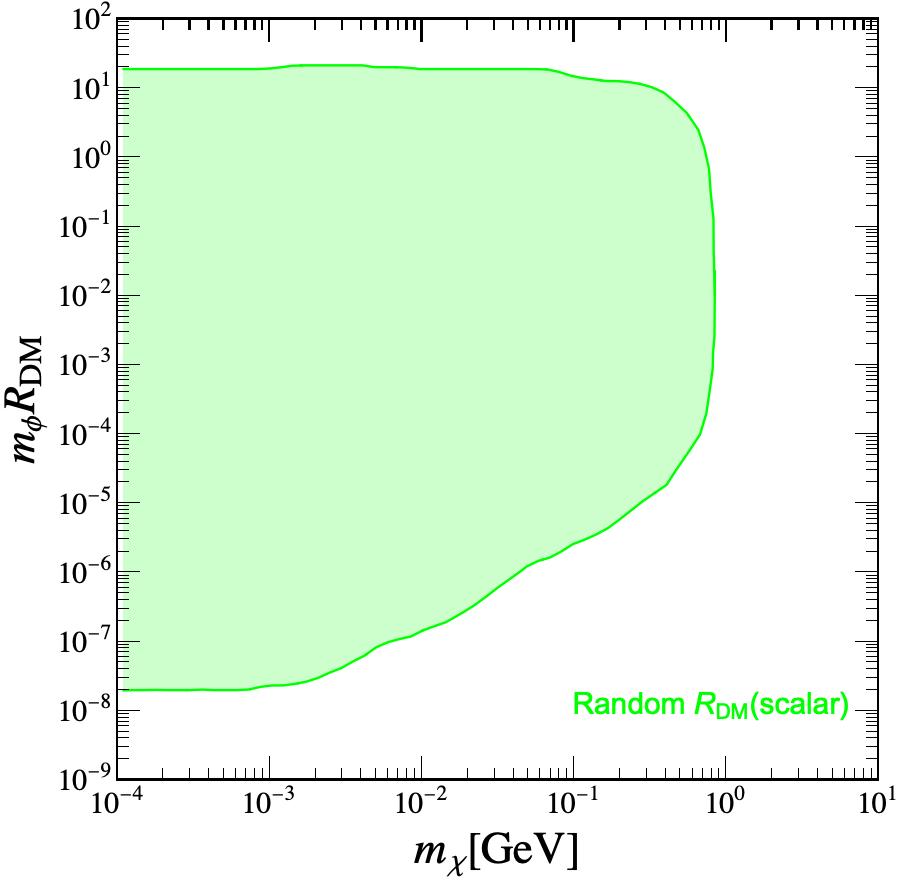}	
	\includegraphics[width=7cm]{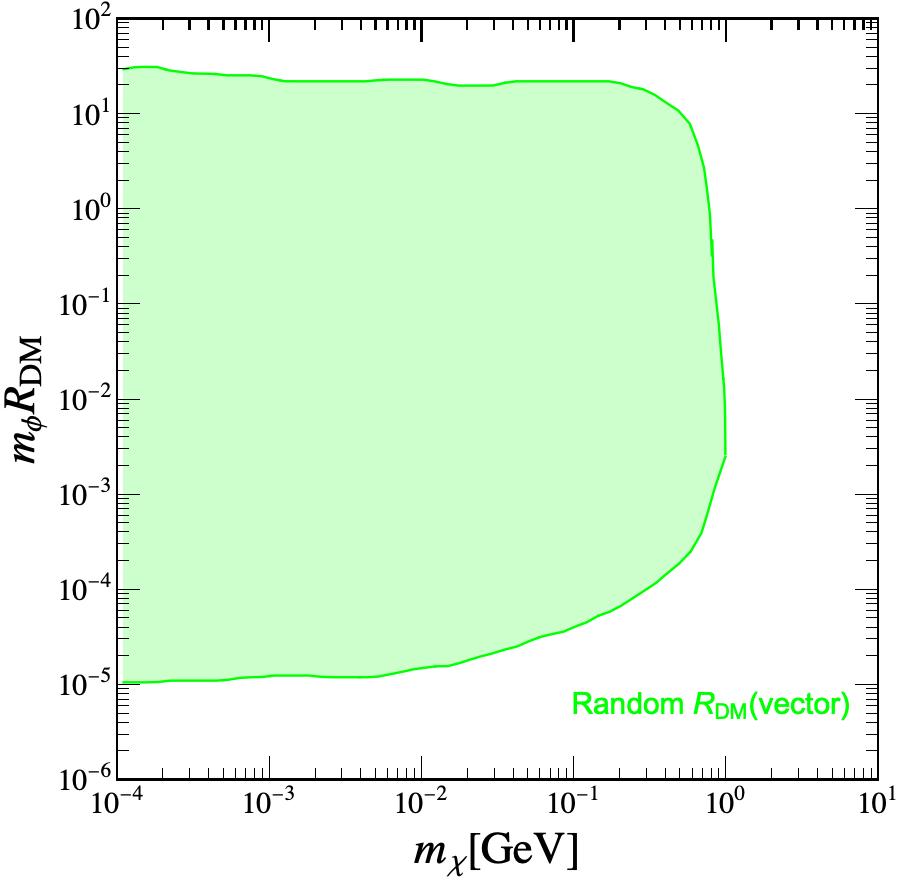}	
  \vspace{-0.4cm}
	\caption{Same as the green region in Fig. (\ref{fig4}), but showing on the plane of DM mass versus the radius-force range ratio $R_{\rm DM}m_{\phi}$. }
	\label{fig7}
\end{figure}

Furthermore, when the radius-force range ratio is varied, we scan over the DM radius  and coupling constants to obtain the light green regions in Fig. \ref{fig4} and  Fig. \ref{fig5}. We see that in this case the excluded parameter space is enlarged. 
 Also, in Fig. \ref{fig6} and  Fig. \ref{fig7} we re-display these constraints on the plane of the coupling constants or the radius-force ratio $R_{\rm DM}m_{\phi}$  versus the DM mass.
 
Finally, we re-do the calculations using other form factors for the puffy DM particle, i.e., the shape of tophat or Gaussian  (shown in the Table I of Ref.~\cite{Chu:2018faw}), which is different from the dipole shape in Eq.(\ref{eq1}) used through out our calculations. We find that all these form factors give the nearly same results. This may be understood as the follows:
As shown in Fig.~1 of Ref. \cite{Chu:2018faw}, the form factor $F_{\rm DM}(qr_{\rm DM})$  is a function of momentum-transfer $q$ and $r_{\rm DM}$ with different distributions. When a momentum transfer is much smaller than $1/r_{\rm DM}$ ($qr_{\rm DM}$ is very small), the internal structure of the DM particle can be hardly measured, namely, $F_{\rm DM}(qr_{\rm DM})\sim 1$,  and thus the effect of form factor is negligible. In the intermediate range of $qr_{\rm DM}$, the value of $F_{\rm DM}(qr_{\rm DM})$ is decreasing and nearly equal for different form factor forms. When the value of $qr_{\rm DM}$ is very big, the scattering cross section is very small. So different form factors yield similar values for the total cross section. 

\section{Conclusion}\label{sec5}
In this work we examined  the direct detection limits on the cosmic ray-boosted puffy DM for a simplified model with a light mediator.
For comparison, we first re-derived the direct detection limits on the cosmic ray-boosted point-like DM.  We displayed the excluded regions in terms of the DM and mediator masses, the coupling constants, and the ratio of radius to force range. 
Our results showed that the momentum-transfer from both the mediator effect and the DM radius effect should be jointly considered.  Compared to point-like DM, the direct detection limits were found to be stronger for the cosmic ray-boosted puffy DM.

\section*{Acknowledgements}
This work was supported by the National
Natural Science Foundation of China (NNSFC) under grant Nos. 11775012, 11821505 and 12075300,
by Peng-Huan-Wu Theoretical Physics Innovation Center (12047503), and by the Key Research Program of the Chinese Academy of Sciences, grant No. XDPB15.

\vspace{0.5cm}
{\bf Note added:} While we were preparing this manuscript (which took a rather long time due to the covid pandemic), a similar work \cite{Alvey:2022pad} appeared in arXiv. That work considered that the momentum-transfer dependence of the cosmic ray-boosted dark matter scattering cross section can be from a light mediator or the size effect of the DM particle. In our work, for the CRPDM scattering cross section  
we considered the momentum-transfer dependence caused jointly by a light mediator and the size effect of the DM particle. As a result, we found that for a small radius-force range  ratio the DM size effect can be neglected, while for a large radius-force range ratio the DM size effect can be dominant.


\bibliographystyle{apsrev}
\bibliography{note}

\end{document}